\documentclass[10pt,leqno]{amsart}
\usepackage[a4paper,margin=1in]{geometry}
\usepackage{graphicx} 
\usepackage{url}
\usepackage{multirow}%
\usepackage{array}%
\renewcommand{\arraystretch}{1.5}%
\usepackage{amsmath,cite,url}
\usepackage{booktabs}%
\usepackage{breqn}
\usepackage[table]{xcolor}%
\usepackage{dblfloatfix}
\newcolumntype{P}[1]{>{\centering\arraybackslash}p{#1}}
\usepackage{hhline}
\usepackage{accents}%
\usepackage{algorithm}%
\usepackage{algorithmicx}%
\usepackage{algpseudocode}%
\usepackage{listings}%
\usepackage{caption}%
\usepackage{subcaption}%
\usepackage[T1]{fontenc}
\usepackage[utf8]{inputenc}
\usepackage{amsmath,amssymb,amsfonts}%
\usepackage{amsthm}%
\usepackage{mathrsfs}%
\usepackage{bm}
\theoremstyle{definition}

\theoremstyle{remark} 
 
\numberwithin{equation}{section}

\begin{document} 
\title[short title]{R\=ag Classification of Tagore Songs using Symbolic Music Notation and Novel Weighted Distance Measures} 
\author{Chandan Misra} 
\address[Chandan Misra]{XIM University, School of Computer Science and Engineering, 752050, 
Bhubaneswar, India} 
\email[Chandan Misra]{chandan@xim.edu.in} 
\author{Swarup Chattopadhyay} 
\address[Swarup Chattopadhyay]{XIM University, School of Computer Science and Engineering, 752050, 
Bhubaneswar, India} 
\email[Swarup Chattopadhyay]{swarupc@xim.edu.in} 
\subjclass[2000]{Primary xxx, yyyy; Secondary xxxx, yyyy} 
\date{June 1, 2017, accepted December 7, 2017.} 
\keywords{Dxxx gxxx, Axxx gxxx} 
\begin{abstract} 
Rabindra Sangeet, the body of songs written and composed by Rabindranath Tagore, occupies a distinctive position in Indian music by combining poetic expression with melodic ideas drawn from Hindustani r\=ags, Bengali folk traditions, tappa, kīrtan, Baul music, and Western tunes. Although many Tagore songs are associated with r\=ag labels provided by Tagore himself or preserved in authoritative notational traditions, r\=ag identification remains challenging because the songs often reflect creative freedom rather than strict adherence to classical r\=ag grammar.

This paper formulates r\=ag identification in Rabindra Sangeet as a supervised classification problem using symbolic music-sheet notations from \textit{Swarabitan}. Since large-scale annotated audio or music datasets for Rabindra Sangeet are not readily available, this study constructs a r\=ag-labelled symbolic dataset from notated Tagore songs. The work investigates Euclidean distance and cosine similarity for r\=ag classification and introduces a weighted Euclidean distance measure that assigns greater importance to notes belonging to characteristic r\=ag sequences such as arohana and avarohana. Applied within a k-nearest-neighbour framework, the proposed measure improves r\=ag classification by better capturing r\=ag-specific melodic identity.
\end{abstract} 
\maketitle 
\section{Introduction}\label{sec:introduction}

Indian music has evolved through a continuous interaction between classical discipline, regional expression, devotional imagination, poetic thought, and cultural practice. The Hindustani classical tradition, with its highly developed concepts of r\=ag and tāla, provides a sophisticated framework for organizing melody, rhythm, mood, and performance. At the same time, Indian musical culture has never remained restricted to the boundaries of formal classical grammar. It has continuously interacted with folk traditions, devotional forms, literary movements, theatre, dance, and regional musical practices.

Within this rich musical landscape, Rabindra Sangeet occupies a distinctive and culturally significant position. Composed by Rabindranath Tagore, the Nobel Laureate poet, composer, philosopher, and artist, Rabindra Sangeet represents one of the most important song traditions of Bengal. These songs form an integral part of Bengali cultural life in India and Bangladesh and continue to hold a central place in artistic, social, educational, and devotional contexts.

Musically, Tagore's songs draw from a wide range of sources. Many compositions reveal the influence of Hindustani classical r\=ags and tāla-based structures, while others reflect the melodic character of tappa, kīrtan, Baul, Bengali folk music, devotional idioms, and Western tunes \cite{mukherjee2017tagore,mukerji2011music,som2017rabindranath}. Tagore did not merely reproduce these traditions in a fixed or mechanical manner. Instead, he adapted and transformed them according to the expressive needs of the text, the emotional situation of the song, and his own artistic vision. As a result, Rabindra Sangeet retains a deep connection with Indian musical traditions while developing an independent musical identity.

Although, the tonal colours of r\=ags often provide the emotional foundation of the songs, the use of r\=ag in Rabindra Sangeet is not always identical to its use in formal Hindustani classical performance. In many cases, Tagore employed r\=ag elements with creative freedom, allowing poetic meaning and emotional expression to guide the musical structure. Some Tagore songs closely follow the characteristic structure and melodic behaviour of a particular r\=ag. Others combine features of two or more r\=ags, while some songs are traditionally associated with a r\=ag but do not strictly follow its grammatical framework \cite{trivedi2021rabindra,geetabitanSangeetchintaSummary}. 

The compositional structure of Rabindra Sangeet further increases the difficulty of r\=ag identification. For a beginner or non-specialist listener, recognizing the r\=ag from a Tagore song is often difficult because the song may not present the r\=ag in a direct or conventional manner. A single composition may contain a long sequence of notes, repeated melodic phrases, expressive variations, and movements that do not always correspond to the strict grammar of a classical r\=ag. Consequently, predicting the r\=ag basis of a Rabindra Sangeet composition is a challenging task. Existing approaches to r\=ag classification, which are often designed for classical music performances, cannot be directly applied to Rabindra Sangeet without careful adaptation. This is because Tagore's songs are highly diverse in melodic construction.

R\=ag identification is important for several computational music tasks, including music learning, music information retrieval, mood-based classification, recommendation, and computational music generation. In the context of Rabindra Sangeet, this task becomes particularly meaningful because many songs are associated with r\=ag labels, yet their melodic structures often differ from the strict grammar of Hindustani classical compositions. Therefore, identifying the underlying r\=ag of a Tagore song requires examining its melodic behaviour, including the use of \textit{\={A}roh}, \textit{Avroh}, dominant notes, omitted notes, and the relative importance of notes within the composition.

However, this diversity does not make the problem unsuitable for supervised learning. On the contrary, Rabindra Sangeet provides an interesting setting for supervised r\=ag classification because many songs are associated with r\=ag labels provided by Tagore himself or preserved in authoritative notational traditions. These r\=ag labels offer a meaningful ground truth for computational analysis. Therefore, the present study formulates r\=ag identification as a supervised classification problem, where labelled Tagore songs are used to train a model that can learn r\=ag-specific melodic behaviour from symbolic note sequences.

Although r\=ag identification can naturally be formulated as a supervised learning problem, the availability of suitable annotated datasets remains a major limitation in the context of Rabindra Sangeet. In many music classification tasks, annotated audio or music datasets \cite{srinivasamurthy2021saraga,gulati2016indian,shankar2024saraga} with reliable class labels are used to train supervised models \cite{chowdhuri2019phononet,shah2021raga,gulati2016time,kirthika2012review}. However, such large-scale labelled datasets are not readily available for Rabindra Sangeet. This creates a significant barrier to applying existing data-driven r\=ag classification techniques directly to Tagore songs.

To address this limitation, the present work relies on symbolic music-sheet notations rather than annotated audio recordings. The primary source of these notations is \textit{Swarabitan}, the collection of notated songs written and composed by Rabindranath Tagore. \textit{Swarabitan} contains a large body of Rabindra Sangeet compositions along with their musical notations, and therefore provides a structured symbolic representation of the melodic content of Tagore songs.

These notated compositions are useful for computational analysis because the melodic content of each song can be converted into note sequences and note-frequency distributions. Since many of these compositions are associated with r\=ag labels provided by Tagore himself or preserved in authoritative notational traditions, they can be used to construct a supervised learning dataset. In this way, the absence of a ready-made annotated audio dataset is addressed by preparing a r\=ag-labelled symbolic dataset from notated Rabindra Sangeet compositions.

In this work, we prepare a supervised symbolic dataset of $1000$ r\=ag-labelled Tagore songs from \textit{Swarabitan}. Each sample in the dataset corresponds to the note-frequency distribution of a composition and is associated with its appropriate r\=ag label. The construction of this dataset is an important contribution of the work, as it provides a structured ground truth for supervised r\=ag classification in Rabindra Sangeet.

A natural approach to r\=ag identification is to compare a given composition with other compositions whose r\=ag labels are already known. If two compositions are musically similar, they may be expected to share similar r\=ag characteristics. Therefore, an unknown or test composition can be assigned the r\=ag label of its nearest labelled compositions using a distance or similarity-based classifier. In this direction, Euclidean distance and cosine similarity provide simple and widely used measures for quantifying similarity between compositions represented through note-frequency features.

However, standard distance and similarity measures may not be sufficient for Rabindra Sangeet. Preliminary observations show that Euclidean distance and cosine similarity can sometimes produce contradictory or misleading results. In particular, they may assign high similarity or low distance not only to compositions belonging to the same r\=ag, but also to compositions belonging to different r\=ags. This happens because small variations in individual note frequencies may strongly affect the similarity score, while the musically important role of characteristic notes may not be adequately captured. Thus, two compositions may appear numerically similar even when their r\=ag identities are different.

To address this limitation, the present study introduces a r\={a}g-aware weighted Euclidean distance measure. The main idea is to assign greater importance to notes that belong to the prescribed \textit{\={A}roh} and \textit{Avroh} of a r\={a}g, and comparatively lower importance to other notes. This weighting allows the distance measure to better reflect r\={a}g-specific melodic identity rather than treating all note-frequency differences equally. When used with a $k$-nearest-neighbour classifier, the weighted distance measure helps improve the distinction between compositions belonging to similar and dissimilar r\={a}gs.

\section{Dataset}
\label{sec:dataset}

\subsection{Dataset Generation}
\label{sec:dataset-generation}

As mentioned earlier, Tagore's complete collection of songs consists of approximately $2200$ compositions. This collection of songs is known as \textit{Geetobitan} (\textit{The Garden of Songs}), while their musical notations are published separately in \textit{Swarabitan} (\textit{The Garden of Notes}). \textit{Swarabitan}, a book series comprising around $60$ volumes, contains the notations of songs from \textit{Geetobitan} written in the Akarmatrik notation system.

For the present study, the dataset was created by randomly selecting $1000$ compositions from \textit{Swarabitan}. The notes of each selected composition were manually stored in a CSV file along with the corresponding r\={a}g label. Therefore, each sample in the dataset represents a single composition from \textit{Swarabitan}.

Since Rabindrasangeet is based on the R\={a}gs of Hindustani Sangeet (with a little blend of folk genres like Baul and Kirtan), the notes of the compositions span over three octaves or \textit{Saptak}s, namely the middle or \textit{Madhya}, upper or \textit{Taar}, and lower octave or \textit{Mandra Saptak}. The r\={a}gs also follow the same \textit{\={A}roh} and \textit{Avroh} notes of Hindutani Sangeet which depict the sequence of permissible notes in ascending and descending pattern respectively. Additionally, they aid in describing the rag's mood. Each octave consists of $12$ notes which generates $36$ probable notes for each composition. We have indexed each note with an integer starting from $1$ and ending at $36$ as given in Table \ref{tab:notes-indices}.

\begin{table}[!htb]
     \centering 
     \setlength{\tabcolsep}{2.8pt}
    \renewcommand{\arraystretch}{1.0}
    \begin{tabular}{|c|c|c|c|c|c|c|}
        \hline
        \textbf{Note} & \begin{tabular}{@{}c@{}}Shadaj \\ ($\underaccent{\dot}{S}$, $S$, $\dot{S}$)\end{tabular} & \begin{tabular}{@{}c@{}}Komal \\ Rishabh \\ ($\underaccent{\dot}{r}$, $r$, $\dot{r}$)\end{tabular} & \begin{tabular}{@{}c@{}}Suddha \\ Rishabh \\ ($\underaccent{\dot}{R}$, $R$, $\dot{R}$)\end{tabular} & \begin{tabular}{@{}c@{}}Komal \\ Gandhar \\ ($\underaccent{\dot}{g}$, $g$, $\dot{g}$)\end{tabular} & \begin{tabular}{@{}c@{}}Suddha \\ Gandhar \\ ($\underaccent{\dot}{G}$, $G$, $\dot{G}$)\end{tabular} & \begin{tabular}{@{}c@{}}Madhyam \\ ($\underaccent{\dot}{M}$, $M$, $\dot{M}$)\end{tabular} \\
        \hline
        \textbf{Indices} & 1, 13, 25 & 2, 14, 26 & 3, 15, 27 & 4, 16, 28 & 5, 17, 29 & 6, 18, 30 \\
        \hline
        \textbf{Note} & \begin{tabular}{@{}c@{}}Tivr \\ Madhyam \\ ($\underaccent{\dot}{m}$, $m$, $\dot{m}$)\end{tabular} & \begin{tabular}{@{}c@{}}Pancham \\ ($\underaccent{\dot}{P}$, $P$, $\dot{P}$)\end{tabular} & \begin{tabular}{@{}c@{}}Komal \\ Dhaivat \\ ($\underaccent{\dot}{d}$, $d$, $\dot{d}$)\end{tabular} & \begin{tabular}{@{}c@{}}Suddha \\ Dhaivat \\ ($\underaccent{\dot}{D}$, $D$, $\dot{D}$)\end{tabular} & \begin{tabular}{@{}c@{}}Komal \\ Nishad \\ ($\underaccent{\dot}{n}$, $n$, $\dot{n}$)\end{tabular} & \begin{tabular}{@{}c@{}}Suddha \\ Nishad \\ ($\underaccent{\dot}{N}$, $N$, $\dot{N}$)\end{tabular} \\
        \hline
        \textbf{Indices} & 7, 19, 31 & 8, 20, 32 & 9, 21, 33 & 10, 22, 34 & 11, 23, 35 & 12, 24, 36 \\
        \hline
    \end{tabular}  
    \caption{Notes and indices of lower octave or \textit{Mandra Saptak} (indices 1 to 12), Middle octave or \textit{Madhya Saptak} (indices 13 to 24), and upper octave or \textit{Taar Saptak} (indices 25 to 36)}
    \label{tab:notes-indices}
\end{table}

\begin{figure}[!htb]
     \centering
     \vspace{5mm}
     \begin{subfigure}[b]{0.4\textwidth}
         \centering
         \includegraphics[width=\textwidth]{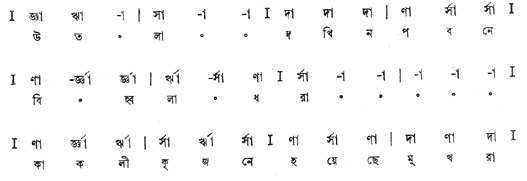}
         \caption{Actual Composition}
         \label{fig:actual-composition}
     \end{subfigure}
     \vspace{5mm}
     \begin{subfigure}[b]{0.5\textwidth}
            {
            \centering
            \small
            \begin{tabular}{|c|c|c|c|c|c|c|c|c|c|c|c|}
                \hline
                16 & 14 & 14 & 13 & 13 & 13 & 21 & 21 & 21 & 23 & 25 & 25 \\
                \hline
                23 & 28 & 28 & 26 & 25 & 23 & 25 & 25 & 25 & 25 & 25 & 25 \\
                \hline
                23 & 28 & 26 & 25 & 26 & 25 & 23 & 25 & 23 & 21 & 23 & 21 \\
                \hline
             \end{tabular}\par             
             }
         \caption{Mapping from actual note to its corresponding integer value}
         \label{fig:mapping}
     \end{subfigure}
     \begin{subfigure}[b]{0.5\textwidth}
            {
            \centering
            \small
            \begin{tabular}{|c|c|c|c|c|c|c|c|c|c|c|c|c|}
                \hline
                \textbf{Note Index} & 1 & 2 & 3 & 4 & 5 & 6 & 7 & 8 & 9 & 10 & 11 & 12 \\
                \hline
                \textbf{Frequency} & 0 & 0 & 0 & 0 & 0 & 0 & 0 & 0 & 0 & 0 & 0 & 0 \\
                \hline
                \textbf{Note Index} & 13 & 14 & 15 & 16 & 17 & 18 & 19 & 20 & 21 & 22 & 23 & 24 \\
                \hline
                \textbf{Frequency} & 3 & 2 & 0 & 1 & 0 & 0 & 0 & 0 & 5 & 0 & 7 & 0 \\
                \hline
                \textbf{Note Index} & 25 & 26 & 27 & 28 & 29 & 30 & 31 & 32 & 33 & 34 & 35 & 36 \\
                \hline
                \textbf{Frequency} & 12 & 3 & 0 & 3 & 0 & 0 & 0 & 0 & 0 & 0 & 0 & 0 \\
                \hline
             \end{tabular}\par             
             }
         \caption{Frequency table of individual notes}
         \label{fig:frequency}
     \end{subfigure}
     
        \caption{The process of note mapping and generating its frequency table for a composition (line no. 19 - 21) titled \textit{kahar galay parabi} belonging to \textit{Prem} parjaay and \textit{Bhairavi} r\={a}g.}
        \label{fig:map-freq-generation}
\end{figure}

Once such mapping is established we generate the note table for each such 1000 compositions. To create the final dataset we generate the frequency distribution of each composition which serves as 36 independent variables and obtain the r\={a}g of the composition for the dependent variable. Figure \ref{fig:map-freq-generation} shows the overall process of mapping a composition to a frequency table.

\subsection{Summary of the Dataset}
\label{sec:summary}

As previously mentioned the dataset consists of 36 features corresponding to 36 notes spanning across three octaves for each sample and there are $1000$ such samples have been considered for our experimental evaluation. We labeled each such sample with the r\={a}g of the composition which have also been extracted manually from the book while creating the dataset and can be considered as a ground truth. There are $239$ unique r\={a}gs identified in the entire dataset and more than 50\% samples ($545$ compositions)  belong to $227$ least frequent r\={a}gs having number of compositions less than $20$. Table 1 shows the frequencies of unique r\={a}gs with increasing number of compositions. In other words Table 1 answers the query \textit{How many unique r\={a}gs are there having $n$ number of compositions?}, for example. 

\begin{table}[!ht]
    \scriptsize
    \begin{subtable}{.46\linewidth}
      \centering
        \begin{tabular}{|c|c|c|}
        \hline
        \textbf{\begin{tabular}{@{}c@{}}\textbf{Number of} \\ \textbf{Compositions}\end{tabular}} & \textbf{\begin{tabular}{@{}c@{}}\textbf{Frequency of } \\ \textbf{Unique R\={a}gs}\end{tabular}}  & \textbf{\begin{tabular}{@{}c@{}}\textbf{Total No. } \\ \textbf{of Compositions}\end{tabular}} \\
        \hline
        1 & 135 & 135 \\
        \hline
        2 & 33 & 66 \\
        \hline
        3 & 15 & 45 \\
        \hline
        4 & 14 & 56 \\
        \hline
        5 & 7 & 35 \\
        \hline
        6 & 9 & 54 \\
        \hline
        7 & 3 & 21 \\
        \hline
        8 & 2 & 16 \\
        \hline
        9 & 3 & 27 \\
        \hline
        10 & 1 & 10 \\
        \hline
        13 & 1 & 13 \\
        \hline
        15 & 1 & 15 \\
        \hline
        16 & 1 & 16 \\
        \hline
        17 & 1 & 17 \\
        \hline
        19 & 1 & 19 \\
        \hline
        \textbf{Total} & 227 & 545\\
        \hline
    \end{tabular}
    \caption{}
    \label{tab:frequency}
    \end{subtable}%
    \hfill
    \begin{subtable}{.54\linewidth}
      \centering
        \begin{tabular}{|c|c|c|c|}
        \hline
        \textbf{R\={a}g} & \textbf{Abbr.} & \textbf{Thaat} & \textbf{Frequency} \\
        \hline
        Bhairavi & $bh$ & Bhairavi & 100 \\
        \hline
        Bihag & $bi$ & Bilaval & 49 \\
        \hline
        Kirtan & $kr$ & NA & 45 \\
        \hline
        Khamaj & $kh$ & Khamaj & 37 \\
        \hline
        Desh & $de$ & Khamaj & 35 \\
        \hline
        Pilu & $pl$ & Kafi & 33 \\
        \hline
        Baul & $bl$ & NA & 33 \\
        \hline
        Kafi & $kf$ & Kafi & 31 \\
        \hline
        Yaman-Kalyan & $yk$ & Kalyan & 27 \\
        \hline
        Sahana & $sh$ & Kafi & 23 \\
        \hline
        Yaman & $yn$ & Kalyan & 22 \\
        \hline
        Kedara & $kd$ & Kalyan & 20 \\
        \hline
        & & & 455 \\
        \hline
    \end{tabular}
    \caption{}
    \label{tab:raag-thaat}
    \end{subtable} 
    \caption{(a) showing summary to the query \textit{How many unique r\={a}gs are there having $n$ number of compositions?} and (b) showing the $12$ most frequent r\={a}gs in the dataset along with the \textit{thaat} (parent scale according to Pt. Vishnu Narayan Bhatkhande) they belong to.}
\end{table}


While Table~\ref{tab:frequency} summarizes the distribution of the less frequent r\={a}gs in the dataset, Table~\ref{tab:raag-thaat} presents the opposite view by listing the $12$ most frequent r\={a}g labels. This information is important because r\={a}gs with very few compositions may not provide sufficient samples to capture a reliable note-frequency pattern or average distribution for that particular r\={a}g. 

In Table~\ref{tab:raag-thaat}, we also provide the corresponding \textit{Thaat} of each r\={a}g. In Hindustani classical music, a \textit{Thaat} refers to the parent scale or melodic framework used in Pt. Vishnu Narayan Bhatkhande's system for classifying r\={a}gs according to their note structure. Including the \textit{Thaat} information helps the reader understand the broader melodic family to which a r\={a}g belongs. It is also useful for interpreting similarities between r\={a}gs, since r\={a}gs belonging to the same \textit{Thaat} may share similar note structures.

However, the entries \textit{Baul} and \textit{Kirtan} are marked as NA because they are not classified as r\={a}gs under the Bhatkhande \textit{Thaat} system. Rather, they represent folk and devotional musical traditions that occur frequently in Rabindra Sangeet but do not necessarily follow a fixed \={A}roh--Avroh structure associated with a specific Hindustani classical r\={a}g. Therefore, they are retained in the dataset as musical-category labels rather than being assigned to a particular \textit{Thaat}.

Thus, the dataset can be viewed as a note-frequency-based representation of individual compositions. Each sample represents the distribution of notes in a composition and is associated with its corresponding r\={a}g or musical-category label.

\section{Notations Used in the Present Work}
\label{sec:notations}

Let $\mathcal{N}$ denote the set of musical notes used in this work, as listed in Table~\ref{tab:notes-indices}. Thus,

\begin{dmath}
\mathcal{N} = \Bigl\{ \underaccent{\dot}{S}, S, \dot{S}, \underaccent{\dot}{r}, r, \dot{r}, \underaccent{\dot}{R}, R, \dot{R}, \underaccent{\dot}{g}, g, \dot{g}, 
\underaccent{\dot}{G}, G, \dot{G}, \underaccent{\dot}{M}, M, \dot{M}, \underaccent{\dot}{m}, m, \dot{m}, \underaccent{\dot}{P}, P, \dot{P}, 
\underaccent{\dot}{d}, d, \dot{d}, \underaccent{\dot}{D}, D, \dot{D}, \underaccent{\dot}{n}, n, \dot{n}, \underaccent{\dot}{N}, N, \dot{N} \Bigr\}.
\end{dmath}

Similarly, let $\mathcal{R}$ denote the set of r\={a}gs considered in this paper. Therefore,

\begin{dmath}
\mathcal{R} = \Bigl\{ bh, bi, kr, kh, de, pl, bl, kf, yk, sh, yn, kd \Bigr\}.
\end{dmath}

The $i$-th sample composition belonging to a particular r\={a}g $\rho$ is denoted by $comp_{\rho}^{i}$, where $\rho \in \mathcal{R}$. For example, the compositions belonging to r\={a}g \textit{Bhairavi} can be denoted by $comp_{bh}^{i}$, where $i \in \{1,2,\dots,100\}$, since the dataset contains $100$ compositions of r\={a}g \textit{Bhairavi}, as shown in Table~\ref{tab:raag-thaat}.

Let $\mathcal{I}$ denote the set of note indices corresponding to the notes listed in Table~\ref{tab:notes-indices}. This set is defined as

\begin{dmath}
\mathcal{I} = \Bigl\{ 1, 2, \dots, 35, 36 \Bigr\}.
\end{dmath}

For a given composition $comp_{\rho}^{i}$ belonging to r\={a}g $\rho$, the set of note indices having non-zero frequency is denoted by $\mathcal{I}_{comp_{\rho}^{i}}$, where

\begin{dmath}
\mathcal{I}_{comp_{\rho}^{i}} \subset \mathcal{I}.
\end{dmath}

For instance, the set of non-zero note indices for the first \textit{Bhairavi} composition in the dataset is given by

\begin{dmath}
\mathcal{I}_{comp_{bh}^{1}} = \{6, 8, 9, 11, 13, 14, 15, 16, 18, 20, 21, 23\}.
\end{dmath}

We further categorize each composition belonging to a r\={a}g $\rho$ as either pure or non-pure using the subscripts $(p)$ and $(np)$, respectively. This categorization is based on the extent to which the composition follows the prescribed \textit{aroh} and \textit{avroh} of the corresponding r\={a}g $\rho$. Accordingly, if the $i$-th composition of r\={a}g $\rho$ is pure, it is denoted by $comp_{\rho(p)}^{i}$; otherwise, if the $i$-th composition is non-pure, it is denoted by $comp_{\rho(np)}^{i}$.

\section{Experimental Evaluation}
\label{sec:experimental-evaluation}
In the experimental evaluation, we quantify the similarity between compositions belonging to the same r\={a}g and those belonging to different r\={a}gs. For this purpose, we employ cosine similarity and Euclidean distance as direct and indirect measures of compositional similarity, respectively. We then propose a modified distance measure to capture the similarity between compositions more effectively. The effectiveness of the proposed measure is further justified through its performance with a $k$-nearest-neighbor classifier.

\subsection{Comparison between compositions belonging to same r\=ag}
\label{sec:comp-between-same-raag-compositions}

\begin{figure}[!htb]
     \centering
     \begin{subfigure}[b]{0.25\textwidth}
         \centering
         \includegraphics[width=\textwidth]{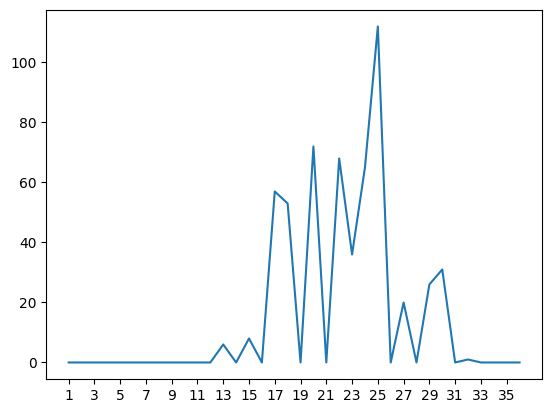}
         \caption{}
         \label{fig:khamaj-dist-1}
     \end{subfigure}
     \begin{subfigure}[b]{0.25\textwidth}
        \centering
        \includegraphics[width=\textwidth]{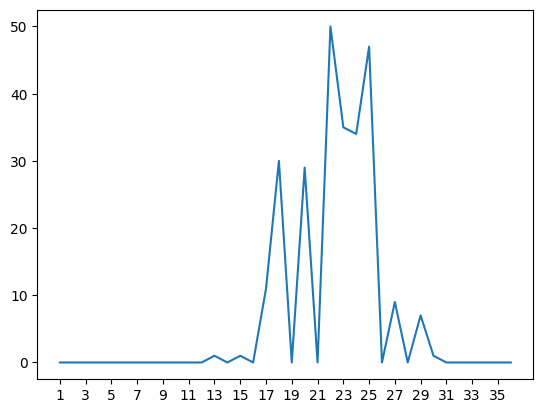}   
         \caption{}
         \label{fig:khamaj-dist-2}
     \end{subfigure}
     \begin{subfigure}[b]{0.25\textwidth}
        \centering
        \includegraphics[width=\textwidth]{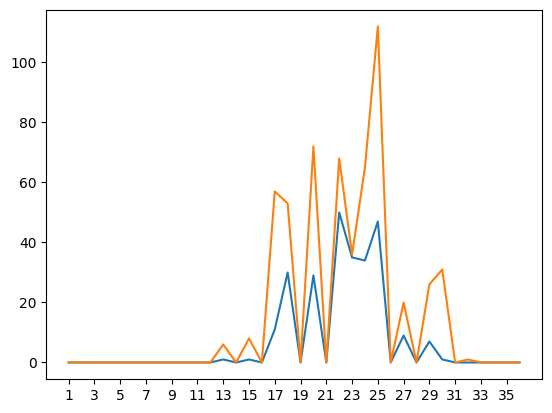}       
        \caption{}
        \label{fig:overplot}
     \end{subfigure}
    \caption{Note Frequency distribution of two pure Khamaj composition \ref{fig:khamaj-dist-1} and \ref{fig:khamaj-dist-2}. X-axis represents note indices as given in Table \ref{tab:notes-indices} and Y-axis represents corresponding note frequencies. Plot \ref{fig:overplot} overplots two distributions which are exactly matching since both belongs to same r\=ag.}
    \label{fig:pure-histogram}
\end{figure}

Compositions belonging to same r\=ag are the ones that follow the same aroh and avroh of the r\=ag. Therefore, the note frequency distribution of the same r\=ag compositions should follow the same silhouette. Table \ref{tab:pure-compositions} and Figure \ref{fig:pure-histogram} shows the note frequency distribution of a pure composition belonging to r\=ag Khamaj.

\begin{table}[!ht]
     \centering 
     \small
    \begin{tabular}{|c|c|c|c|c|c|c|c|c|c|c|c|c|}
        \hline
        \multicolumn{13}{|c|}{Lower Octave or Mandra Saptak} \\
        \hline
        \textbf{Note Index} & \cellcolor{blue!25}1 & 2 & \cellcolor{blue!25}3 & 4 & \cellcolor{blue!25}5 & \cellcolor{blue!25}6 & 7 & \cellcolor{blue!25}8 & 9 & \cellcolor{blue!25}10 & \cellcolor{blue!25}11 & \cellcolor{blue!25}12 \\
        \hline
        $comp_{kh}^{1}$ & 0 & 0 & 0 & 0 & 0 & 0 & 0 & 0 & 0 & 0 & 0 & 0 \\
        \hline
        $comp_{kh}^{5}$ & 0 & 0 & 0 & 0 & 0 & 0 & 0 & 0 & 0 & 0 & 0 & 0 \\
        \hline
        \multicolumn{13}{|c|}{Middle Octave or Madhya Saptak} \\
        \hline
        \textbf{Note Index} & \cellcolor{blue!25}13 & 14 & \cellcolor{blue!25}15 & 16 & \cellcolor{blue!25}17 & \cellcolor{blue!25}18 & 19 & \cellcolor{blue!25}20 & 21 & \cellcolor{blue!25}22 & \cellcolor{blue!25}23 & \cellcolor{blue!25}24 \\
        \hline
        $comp_{kh}^{1}$ & 1 & 0 & 1 & 0 & 11 & 30 & 0 & 29 & 0 & 50 & 35 & 34 \\
        \hline
         $comp_{kh}^{5}$ & 6 & 0 & 8 & 0 & 57 & 53 & 0 & 72 & 0 & 68 & 36 & 65 \\
        \hline
        \multicolumn{13}{|c|}{Upper Octave or Taar Saptak} \\
        \hline
        \textbf{Note Index} & \cellcolor{blue!25}25 & 26 & \cellcolor{blue!25}27 & 28 & \cellcolor{blue!25}29 & \cellcolor{blue!25}30 & 31 & \cellcolor{blue!25}32 & 33 & \cellcolor{blue!25}34 & \cellcolor{blue!25}35 & \cellcolor{blue!25}36 \\
        \hline
        $comp_{kh}^{1}$ & 47 & 0 & 9 & 0 & 7 & 1 & 0 & 0 & 0 & 0 & 0 & 0 \\
        \hline
         $comp_{kh}^{5}$ & 112 & 0 & 20 & 0 & 26 & 31 & 0 & 1 & 0 & 0 & 0 & 0 \\
        \hline
    \end{tabular}  
    \caption{Example of two pure compositions of r\=ag Khamaj. The aroh and avroh notes of r\=ag Khamaj are S, R, G, M, P, D, n, and N and are colored as blue.}
    \label{tab:pure-compositions}
\end{table}

As mentioned earlier, cosine similarity and Euclidean distance are used to analyze the similarity between compositions belonging to the same r\={a}g. Since compositions of the same r\={a}g are expected to exhibit similar note-frequency patterns, they should ideally produce higher cosine similarity values and lower Euclidean distance values.

\begin{figure}[!ht]
    \centering
    \includegraphics[width=0.75\textwidth]{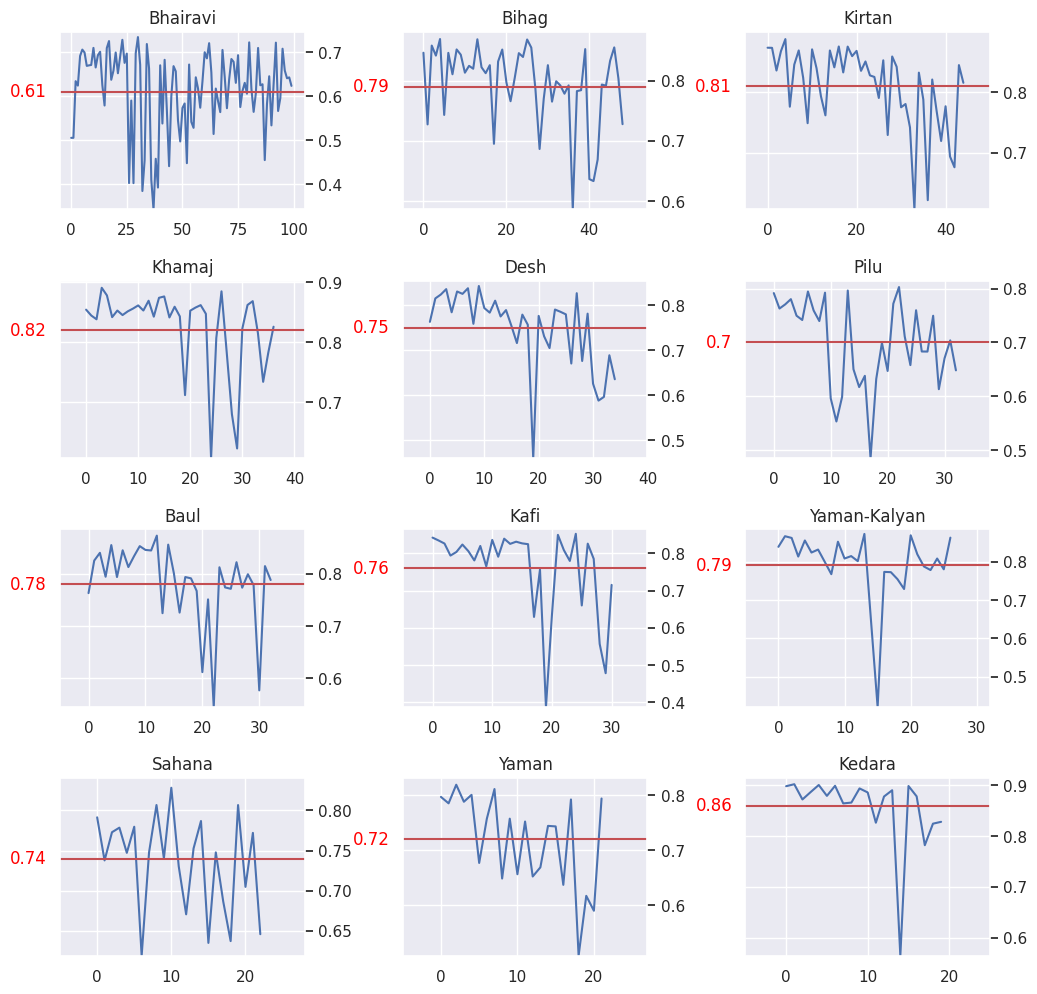}
    \caption{Cosine Similarities between compositions belonging to r\={a}gs in row-major order \textit{Bhairavi}, \textit{Bihag}, \textit{Kirtan}, \textit{Khamaj}, \textit{Desh}, \textit{Pilu}, \textit{Baul}, \textit{Kafi}, \textit{Yaman Kalyan}, \textit{Sahana}, \textit{Yaman}, \textit{Kedara}. The x-axis and y-axis of each sub-plot of the grid correspond to the number of compositions and mean cosine similarity value respectively.}
    \label{fig:unsorted-cosine-sim-same-raag-mean}
\end{figure}

Since cosine similarity ranges from $0$ to $1$, we compute, for each composition, its mean cosine similarity with all other compositions belonging to the same r\={a}g. These mean similarity values are plotted in Figure~\ref{fig:unsorted-cosine-sim-same-raag-mean} to visualize the degree of similarity among compositions of the same r\={a}g.


\begin{figure}[!ht]
    \centering
    \includegraphics[width=0.75\textwidth]{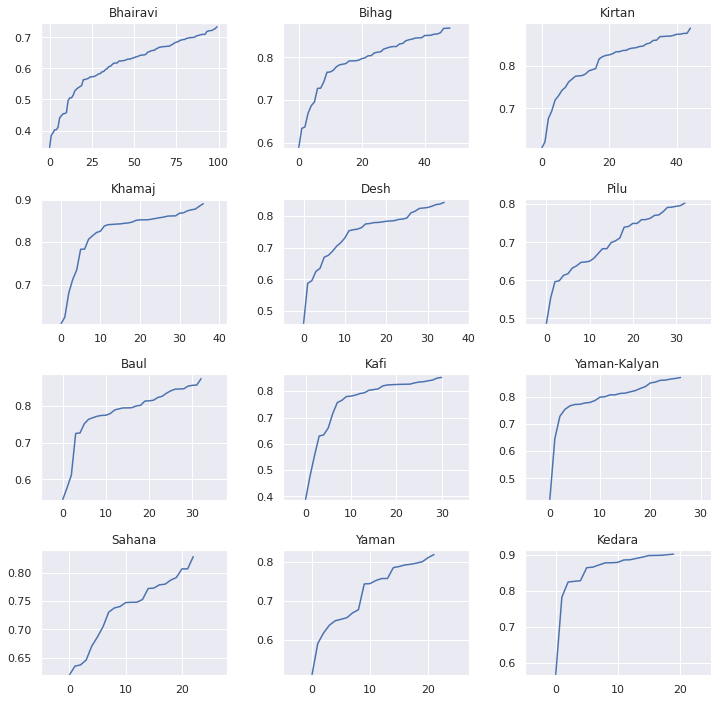}
    \caption{Sorted Mean Cosine Similarities between compositions belonging to r\={a}gs in row-major order \textit{Bhairavi}, \textit{Bihag}, \textit{Kirtan}, \textit{Khamaj}, \textit{Desh}, \textit{Pilu}, \textit{Baul}, \textit{Kafi}, \textit{Yaman Kalyan}, \textit{Sahana}, \textit{Yaman}, \textit{Kedara}. The x-axis and y-axis of each sub-plot of the grid correspond to the number of compositions and mean cosine similarity value respectively.}
    \label{fig:cosine-sim-same-raag}
\end{figure}

Each subfigure in Figure~\ref{fig:unsorted-cosine-sim-same-raag-mean} is overplotted with a red horizontal line representing the mean cosine similarity value. These mean values indicate that, in general, compositions belonging to the same r\={a}g exhibit high similarity among themselves, thereby supporting the observation stated earlier. Except for the compositions belonging to r\={a}g \textit{Bhairavi}, most of the r\={a}g groups show relatively high intra-r\={a}g similarity.

Another way to interpret the similarity among compositions of the same r\={a}g is to examine the distribution of their mean similarity scores. Ideally, only a small number of compositions should have low similarity values, while the majority should exhibit high similarity values. To visualize this pattern more clearly, the mean similarity scores are sorted and plotted in Figure~\ref{fig:cosine-sim-same-raag}. An ideal similarity curve is expected to begin at a relatively high value, rise sharply, and then stabilize close to one. Such behaviour can be observed in the plots corresponding to r\={a}gs \textit{Bihag}, \textit{Kirtan}, \textit{Khamaj}, \textit{Kedara}, and \textit{Yaman Kalyan}.

Deviation from this ideal behavior infers low similarity value and produces plots that gradually increase and exhibits a large number of lower similarity values, r\={a}g Bhairavi and Yaman for example. 

There are two possible reasons for such deviation. First, some compositions may contain notes that do not belong to the prescribed \textit{\={A}roh} and \textit{Avroh} of the corresponding r\={a}g. Second, even when two compositions belong to the same r\={a}g, their notes may occur in different octaves, resulting in little or no overlap in their note-index positions. To illustrate these two cases, we present two representative examples from the dataset.

\begin{enumerate}

\item To support the first case, we consider one pure composition belonging to r\={a}g \textit{Khamaj}, one non-pure composition belonging to r\={a}g \textit{Khamaj}, and one pure composition belonging to r\={a}g \textit{Bhairavi}. The pure Khamaj composition is denoted by $comp_{kh(p)}^{22}$, the non-pure Khamaj composition is denoted by $comp_{kh(np)}^{35}$, and the pure Bhairavi composition is denoted by $comp_{bh(p)}^{43}$, as shown in Table~\ref{tab:weighted-distance-cross-raag}.

For r\={a}g \textit{Khamaj}, the prescribed note-index set is
\[
\mathcal{I}_{kh}=\{1,3,5,6,8,10,11,12\}.
\]
After extending this set across three octaves, the allowed Khamaj note indices are
\[
\{1,3,5,6,8,10,11,12,13,15,17,18,20,22,23,24,25,27,29,30,32,34,35,36\}.
\]

The composition $comp_{kh(p)}^{22}$ is a pure Khamaj composition because all its non-zero note indices belong to the prescribed Khamaj note-index set. In contrast, $comp_{kh(np)}^{35}$ is a non-pure Khamaj composition because it contains the note at index $19$, highlighted in red, which does not belong to the prescribed \textit{\={A}roh} or \textit{Avroh} note set of r\={a}g \textit{Khamaj}. The third composition, $comp_{bh(p)}^{43}$, is a pure Bhairavi composition.

Using normalized ordinary Euclidean distance on the $36$-dimensional note-frequency vectors, the distance between $comp_{kh(p)}^{22}$ and $comp_{bh(p)}^{43}$ is $0.3099$, whereas the distance between $comp_{kh(p)}^{22}$ and $comp_{kh(np)}^{35}$ is $0.3242$. Thus, normalized ordinary Euclidean distance incorrectly indicates that the pure Khamaj composition is closer to the pure Bhairavi composition than to the non-pure Khamaj composition. This observation motivates the use of a r\={a}g-aware weighted distance measure, discussed later, to better preserve musically meaningful relationships.

\begin{table}[!ht]
 \centering 
 \scriptsize
\begin{tabular}{|c|c|c|c|c|c|c|c|c|c|c|c|c|}
    \hline
    \multicolumn{13}{|c|}{Lower Octave or Mandra Saptak} \\
    \hline
    \textbf{Note Index} 
    & \cellcolor{blue!25}1 & 2 & \cellcolor{blue!25}3 & 4 
    & \cellcolor{blue!25}5 & \cellcolor{blue!25}6 & 7 
    & \cellcolor{blue!25}8 & 9 & \cellcolor{blue!25}10 
    & \cellcolor{blue!25}11 & \cellcolor{blue!25}12 \\
    \hline
    $comp_{kh(p)}^{22}$ 
    & 0 & 0 & 0 & 0 & 0 & 0 & 0 & 0 & 0 & 0 & 0 & 0 \\
    \hline
    $comp_{kh(np)}^{35}$ 
    & 0 & 0 & 0 & 0 & 0 & 0 & 0 & 0 & 0 & 4 & 0 & 0 \\
    \hline
    $comp_{bh(p)}^{43}$ 
    & 0 & 0 & 0 & 0 & 0 & 0 & 0 & 0 & 0 & 0 & 0 & 0 \\
    \hline

    \multicolumn{13}{|c|}{Middle Octave or Madhya Saptak} \\
    \hline
    \textbf{Note Index} 
    & \cellcolor{blue!25}13 & 14 & \cellcolor{blue!25}15 & 16 
    & \cellcolor{blue!25}17 & \cellcolor{blue!25}18 & 19 
    & \cellcolor{blue!25}20 & 21 & \cellcolor{blue!25}22 
    & \cellcolor{blue!25}23 & \cellcolor{blue!25}24 \\
    \hline
    $comp_{kh(p)}^{22}$ 
    & 17 & 0 & 8 & 0 & 7 & 44 & 0 & 42 & 0 & 63 & 42 & 14 \\
    \hline
    $comp_{kh(np)}^{35}$ 
    & 27 & 0 & 14 & 0 & 67 & 10 & \cellcolor{red!40}9 & 75 & 0 & 28 & 0 & 17 \\
    \hline
    $comp_{bh(p)}^{43}$ 
    & 10 & 16 & 0 & 12 & 0 & 15 & 0 & 21 & 20 & 0 & 17 & 0 \\
    \hline

    \multicolumn{13}{|c|}{Upper Octave or Taar Saptak} \\
    \hline
    \textbf{Note Index} 
    & \cellcolor{blue!25}25 & 26 & \cellcolor{blue!25}27 & 28 
    & \cellcolor{blue!25}29 & \cellcolor{blue!25}30 & 31 
    & \cellcolor{blue!25}32 & 33 & \cellcolor{blue!25}34 
    & \cellcolor{blue!25}35 & \cellcolor{blue!25}36 \\
    \hline
    $comp_{kh(p)}^{22}$ 
    & 55 & 0 & 19 & 0 & 4 & 0 & 0 & 0 & 0 & 0 & 0 & 0 \\
    \hline
    $comp_{kh(np)}^{35}$ 
    & 38 & 0 & 4 & 0 & 0 & 0 & 0 & 0 & 0 & 0 & 0 & 0 \\
    \hline
    $comp_{bh(p)}^{43}$ 
    & 18 & 3 & 0 & 1 & 0 & 0 & 0 & 0 & 0 & 0 & 0 & 0 \\
    \hline

    \multicolumn{13}{|c|}{\textbf{Normalized Ordinary Euclidean Distance}} \\
    \hline
    \multicolumn{7}{|c|}{$comp_{kh(p)}^{22}$ vs. $comp_{bh(p)}^{43}$} 
    & \multicolumn{6}{|c|}{0.3099} \\
    \hline
    \multicolumn{7}{|c|}{$comp_{kh(p)}^{22}$ vs. $comp_{kh(np)}^{35}$} 
    & \multicolumn{6}{|c|}{0.3242} \\
    \hline
\end{tabular}  

\caption{Comparison of a pure Khamaj composition $comp_{kh(p)}^{22}$, a non-pure Khamaj composition $comp_{kh(np)}^{35}$, and a pure Bhairavi composition $comp_{bh(p)}^{43}$. The prescribed Khamaj note indices are highlighted in blue, while the deviated note index in the non-pure Khamaj composition is highlighted in red. Normalized ordinary Euclidean distance gives a misleading ordering by placing the pure Khamaj composition closer to the pure Bhairavi composition than to the non-pure Khamaj composition.}
\label{tab:weighted-distance-cross-raag}
\end{table}

\item To support the second case, we consider two compositions belonging to r\={a}g \textit{Bhairavi}. Both compositions are non-pure, i.e., they do not strictly follow the prescribed \textit{\={A}roh} and \textit{Avroh} pattern of r\={a}g \textit{Bhairavi}. However, their notes are distributed across different octaves in a non-overlapping manner. That is,
\[
    \mathcal{I}_{comp_{bh(np)}^{312}}
    \cap
    \mathcal{I}_{comp_{bh(np)}^{415}}
    =
    \emptyset .
\]

Since both compositions belong to r\={a}g \textit{Bhairavi}, their similarity is expected to be high from a r\={a}g-level perspective. However, due to the non-overlapping nature of their note-index positions across the three octaves, the cosine similarity becomes zero, as shown in Table~\ref{tab:non-overlapped-notes-composition-cos-sim}. This example further indicates that cosine similarity may fail to capture r\={a}g-level similarity when octave-wise note positions differ significantly.

\begin{table}[!ht]
 \centering 
 \small
\begin{tabular}{|c|c|c|c|c|c|c|c|c|c|c|c|c|}
    \hline
    \multicolumn{13}{|c|}{Lower Octave or Mandra Saptak} \\
    \hline
    \textbf{Note Index} & 1 & 2 & 3 & 4 & 5 & 6 & 7 & 8 & 9 & 10 & 11 & 12 \\
    \hline
    $comp_{bh(np)}^{312}$ 
    & 0 & 0 & 0 & 0 & 0 & 0 & 0 & 0 & 0 & 0 & 0 & 0 \\
    \hline
    $comp_{bh(np)}^{415}$ 
    & 0 & 0 & 0 & 0 & 0 & \cellcolor{blue!25}4 & 0 & \cellcolor{blue!25}5 & \cellcolor{blue!25}25 & 0 & \cellcolor{blue!25}36 & \cellcolor{blue!25}1 \\
    \hline

    \multicolumn{13}{|c|}{Middle Octave or Madhya Saptak} \\
    \hline
    \textbf{Note Index} & 13 & 14 & 15 & 16 & 17 & 18 & 19 & 20 & 21 & 22 & 23 & 24 \\
    \hline
    $comp_{bh(np)}^{312}$ 
    & 0 & 0 & 0 & 0 & 0 & 0 & 0 & 0 & \cellcolor{blue!25}4 & 0 & \cellcolor{blue!25}10 & \cellcolor{blue!25}6 \\
    \hline
    $comp_{bh(np)}^{415}$ 
    & \cellcolor{blue!25}68 & \cellcolor{blue!25}29 & \cellcolor{blue!25}3 & \cellcolor{blue!25}31 & 0 & \cellcolor{blue!25}9 & 0 & 0 & 0 & 0 & 0 & 0 \\
    \hline

    \multicolumn{13}{|c|}{Upper Octave or Taar Saptak} \\
    \hline
    \textbf{Note Index} & 25 & 26 & 27 & 28 & 29 & 30 & 31 & 32 & 33 & 34 & 35 & 36 \\
    \hline
    $comp_{bh(np)}^{312}$ 
    & \cellcolor{blue!25}40 & \cellcolor{blue!25}23 & \cellcolor{blue!25}1 & \cellcolor{blue!25}29 & 0 & \cellcolor{blue!25}2 & 0 & 0 & 0 & 0 & 0 & 0 \\
    \hline
    $comp_{bh(np)}^{415}$ 
    & 0 & 0 & 0 & 0 & 0 & 0 & 0 & 0 & 0 & 0 & 0 & 0 \\
    \hline

    \multicolumn{13}{|c|}{\textbf{Cosine Similarity Score}} \\
    \hline
    \multicolumn{7}{|c|}{$comp_{bh(np)}^{312}$ vs. $comp_{bh(np)}^{415}$} 
    & \multicolumn{6}{|c|}{0.0} \\
    \hline
\end{tabular}  
\caption{Two non-pure compositions, $comp_{bh(np)}^{312}$ and $comp_{bh(np)}^{415}$, belonging to r\={a}g \textit{Bhairavi}. Although both compositions belong to the same r\={a}g, their non-zero note indices occur in non-overlapping octave-wise positions, as highlighted in blue. Consequently, the cosine similarity between them becomes zero.}
\label{tab:non-overlapped-notes-composition-cos-sim}
\end{table}
\end{enumerate}

\subsection{Comparison between compositions belonging to different r\=ags}
\label{sec:comp-between-diff-raag-compositions}

A similar issue arises when computing the similarity between compositions belonging to different r\={a}gs. Since r\={a}gs belonging to the same \textit{Thaat} may exhibit similar melodic structures, we selected five r\={a}gs, namely \textit{Bhairavi}, \textit{Bihag}, \textit{Khamaj}, \textit{Kafi}, and \textit{Kedara}, each belonging to a different \textit{Thaat}, as shown in Table~\ref{tab:raag-thaat}. 

\begin{figure*}[!ht]
    \centering
    \includegraphics[width=\textwidth]{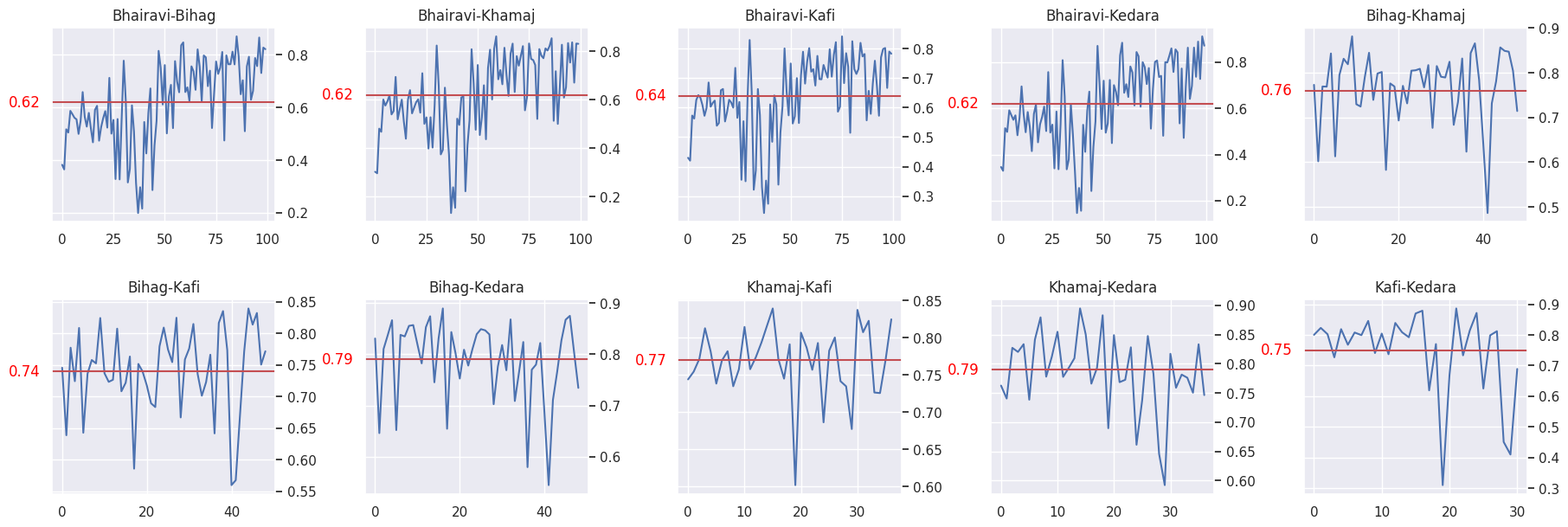}
    \caption{Mean cosine similarity scores between compositions belonging to different r\={a}gs. The selected r\={a}gs are Bhairavi, Bihag, Khamaj, Kafi, and Kedara, each shown in pairwise comparison. The x-axis represents the composition index, while the y-axis represents the mean cosine similarity with compositions of the other r\={a}g. The horizontal red line shows the average similarity score for each r\={a}g pair.}
    \label{fig:mean-unsorted-cos-sim-different-raag}
\end{figure*}

Similar to the experiment conducted for compositions belonging to the same r\={a}g, we visualize the similarity between compositions belonging to different r\={a}gs by computing the mean cosine similarity of each composition with all compositions of another r\={a}g. The resulting plots are shown in Figure~\ref{fig:mean-unsorted-cos-sim-different-raag}. Each subfigure in Figure~\ref{fig:mean-unsorted-cos-sim-different-raag} is overplotted with a red horizontal line, which represents the mean similarity value for the corresponding pair of r\={a}gs.

The mean similarity score indicated by the red horizontal line shows that most of the similarity scores between compositions belonging to different r\={a}gs lie above $0.62$, which is a considerably high value in terms of cosine similarity. For example, the mean similarity score between r\={a}g \textit{Bihag} and r\={a}g \textit{Kedara} is $0.79$. Such a high similarity value between compositions of two different r\={a}gs is not generally expected from a musical point of view.

This behaviour can be further understood through a representative example involving pure compositions. We consider one pure composition from r\={a}g \textit{Bhairavi}, namely $comp_{bh(p)}^{43}$, and three pure compositions from r\={a}g \textit{Khamaj}, namely $comp_{kh(p)}^{17}$, $comp_{kh(p)}^{20}$, and $comp_{kh(p)}^{21}$. The note-frequency distributions of these compositions and the corresponding cosine similarity scores are shown in Table~\ref{tab:simi-different-raag}.

\begin{table}[!ht]
 \centering 
 \small
\begin{tabular}{|c|c|c|c|c|c|c|c|c|c|c|c|c|}
    \hline
    \multicolumn{13}{|c|}{Lower Octave or Mandra Saptak} \\
    \hline
    \textbf{Note Index} & 1 & 2 & 3 & 4 & 5 & 6 & 7 & 8 & 9 & 10 & 11 & 12 \\
    \hline
    $comp_{bh(p)}^{43}$ & 0 & 0 & 0 & 0 & 0 & 0 & 0 & 0 & 0 & 0 & 0 & 0 \\
    \hline
    $comp_{kh(p)}^{17}$ & 0 & 0 & 0 & 0 & 0 & 0 & 0 & 0 & 0 & 0 & 0 & 0 \\
    \hline
    $comp_{kh(p)}^{20}$ & 0 & 0 & 0 & 0 & 0 & 0 & 0 & 0 & 0 & 0 & 0 & 0 \\
    \hline
    $comp_{kh(p)}^{21}$ & 0 & 0 & 0 & 0 & 0 & 0 & 0 & 0 & 0 & 0 & 0 & 0 \\
    \hline

    \multicolumn{13}{|c|}{Middle Octave or Madhya Saptak} \\
    \hline
    \textbf{Note Index} & 13 & 14 & 15 & 16 & 17 & 18 & 19 & 20 & 21 & 22 & 23 & 24 \\
    \hline
    $comp_{bh(p)}^{43}$ & 10 & 16 & 0 & 12 & 0 & 15 & 0 & 21 & 20 & 0 & 17 & 0 \\
    \hline
    $comp_{kh(p)}^{17}$ & 0 & 0 & 2 & 0 & 5 & 13 & 0 & 29 & 0 & 19 & 21 & 2 \\
    \hline
    $comp_{kh(p)}^{20}$ & 0 & 0 & 0 & 0 & 1 & 5 & 0 & 2 & 0 & 1 & 4 & 9 \\
    \hline
    $comp_{kh(p)}^{21}$ & 4 & 0 & 5 & 0 & 31 & 31 & 0 & 47 & 0 & 20 & 14 & 11 \\
    \hline

    \multicolumn{13}{|c|}{Upper Octave or Taar Saptak} \\
    \hline
    \textbf{Note Index} & 25 & 26 & 27 & 28 & 29 & 30 & 31 & 32 & 33 & 34 & 35 & 36 \\
    \hline
    $comp_{bh(p)}^{43}$ & 18 & 3 & 0 & 1 & 0 & 0 & 0 & 0 & 0 & 0 & 0 & 0 \\
    \hline
    $comp_{kh(p)}^{17}$ & 26 & 0 & 6 & 0 & 10 & 4 & 0 & 1 & 0 & 0 & 0 & 0 \\
    \hline
    $comp_{kh(p)}^{20}$ & 10 & 0 & 1 & 0 & 0 & 0 & 0 & 0 & 0 & 0 & 0 & 0 \\
    \hline
    $comp_{kh(p)}^{21}$ & 27 & 0 & 7 & 0 & 0 & 0 & 0 & 0 & 0 & 0 & 0 & 0 \\
    \hline

    \multicolumn{13}{|c|}{\textbf{Cosine Similarity Scores}} \\
    \hline
    \multicolumn{7}{|c|}{$comp_{bh(p)}^{43}$ vs. $comp_{kh(p)}^{17}$} 
    & \multicolumn{6}{|c|}{\cellcolor{red!40}0.6733} \\
    \hline
    \multicolumn{7}{|c|}{$comp_{kh(p)}^{20}$ vs. $comp_{kh(p)}^{21}$} 
    & \multicolumn{6}{|c|}{0.6426} \\
    \hline
\end{tabular}  
\caption{Composition $comp_{bh(p)}^{43}$ is a pure composition belonging to r\={a}g \textit{Bhairavi}, while $comp_{kh(p)}^{17}$, $comp_{kh(p)}^{20}$, and $comp_{kh(p)}^{21}$ are pure compositions belonging to r\={a}g \textit{Khamaj}. The cosine similarity between the pure compositions of two different r\={a}gs, $comp_{bh(p)}^{43}$ and $comp_{kh(p)}^{17}$, is higher than the cosine similarity between two pure compositions of the same r\={a}g, $comp_{kh(p)}^{20}$ and $comp_{kh(p)}^{21}$.}
\label{tab:simi-different-raag}
\end{table}

From Table~\ref{tab:simi-different-raag}, it can be observed that the cosine similarity between two pure compositions belonging to different r\={a}gs, namely $comp_{bh(p)}^{43}$ and $comp_{kh(p)}^{17}$, is $0.6733$. In contrast, the cosine similarity between two pure compositions belonging to the same r\={a}g \textit{Khamaj}, namely $comp_{kh(p)}^{20}$ and $comp_{kh(p)}^{21}$, is $0.6426$. Thus,

\begin{dmath}
\cos\bigl(comp_{bh(p)}^{43}, comp_{kh(p)}^{17}\bigr)
>
\cos\bigl(comp_{kh(p)}^{20}, comp_{kh(p)}^{21}\bigr).
\end{dmath}

This example shows that two pure compositions belonging to different r\={a}gs may exhibit a higher cosine similarity score than two pure compositions belonging to the same r\={a}g. Therefore, quantifying similarity or dissimilarity among r\={a}g-based compositions cannot be reliably achieved using cosine similarity alone. Such observations motivate the need to modify the structure of the dataset and the distance measure, as described in Section~\ref{sec:dataset-modification} and Section~\ref{sec:dis-measure}, respectively.

\subsection{Dataset Modification}
\label{sec:dataset-modification}

In the original representation, each composition is represented using the frequency distribution of $36$ notes, corresponding to $12$ notes across three octaves. However, for r\={a}g-level similarity analysis, the octave position of a note does not necessarily change the identity of the r\={a}g. For example, the occurrence of the same note in the mandra, madhya, or taar saptak contributes to the melodic structure of the same r\={a}g. Therefore, instead of using the complete $36$-dimensional note-frequency vector, we reduce each composition to a $12$-dimensional cumulative note-frequency vector.

Let $\mathcal{N}_{12}$ denote the reduced set of $12$ notes, defined as
\[
\mathcal{N}_{12}
=
\bigl\{
S,r,R,g,G,M,m,P,d,D,n,N
\bigr\}.
\]
The corresponding note-index set is defined as
\[
\mathcal{I}_{12}
=
\{1,2,\dots,12\}.
\]

Let $x = (x_1,x_2,\dots,x_{36})$ be the original $36$-dimensional note-frequency vector of a composition. The corresponding $12$-dimensional cumulative vector $x'=(x'_1,x'_2,\dots,x'_{12})$ is obtained by summing the frequencies of the same note across the three octaves. Thus,
\[
x'_j
=
x_j + x_{j+12} + x_{j+24},
\qquad
j=1,2,\dots,12.
\]

This mapping combines the frequencies of corresponding notes from the lower, middle, and upper octaves. As a result, the comparison between two compositions becomes less sensitive to octave-specific positions and more focused on the overall note usage pattern of the r\={a}g. This modification helps reduce the possibility of obtaining very low or zero similarity scores between compositions of the same r\={a}g merely because their notes occur in different octaves.

Table~\ref{tab:mapping} illustrates this transformation using the composition $comp_{kh}^{1}$.

\begin{table}[!ht]
     \centering 
     \small
    \begin{tabular}{|c|c|c|c|c|c|c|c|c|c|c|c|c|}
        \hline
        \multicolumn{13}{|c|}{Lower Octave or Mandra Saptak} \\
        \hline
        \textbf{Note Index} & 1 & 2 & 3 & 4 & 5 & 6 & 7 & 8 & 9 & 10 & 11 & 12 \\
        \hline
        $comp_{kh}^{1}$ & 0 & 0 & 0 & 0 & 0 & 0 & 0 & 0 & 0 & 0 & 0 & 0 \\
        \hline

        \multicolumn{13}{|c|}{Middle Octave or Madhya Saptak} \\
        \hline
        \textbf{Note Index} & 13 & 14 & 15 & 16 & 17 & 18 & 19 & 20 & 21 & 22 & 23 & 24 \\
        \hline
        $comp_{kh}^{1}$ & 1 & 0 & 1 & 0 & 11 & 30 & 0 & 29 & 0 & 50 & 35 & 34 \\
        \hline

        \multicolumn{13}{|c|}{Upper Octave or Taar Saptak} \\
        \hline
        \textbf{Note Index} & 25 & 26 & 27 & 28 & 29 & 30 & 31 & 32 & 33 & 34 & 35 & 36 \\
        \hline
        $comp_{kh}^{1}$ & 47 & 0 & 9 & 0 & 7 & 1 & 0 & 0 & 0 & 0 & 0 & 0 \\
        \hline

        \multicolumn{13}{|c|}{Aggregated Frequency Distribution across Octaves} \\
        \hline
        \textbf{\begin{tabular}{@{}c@{}}\textbf{Note Index} \\ \textbf{Note}\end{tabular}} 
        & \begin{tabular}{@{}c@{}}1 \\ $S$\end{tabular} 
        & \begin{tabular}{@{}c@{}}2 \\ $r$\end{tabular} 
        & \begin{tabular}{@{}c@{}}3 \\ $R$\end{tabular} 
        & \begin{tabular}{@{}c@{}}4 \\ $g$\end{tabular} 
        & \begin{tabular}{@{}c@{}}5 \\ $G$\end{tabular} 
        & \begin{tabular}{@{}c@{}}6 \\ $M$\end{tabular} 
        & \begin{tabular}{@{}c@{}}7 \\ $m$\end{tabular} 
        & \begin{tabular}{@{}c@{}}8 \\ $P$\end{tabular} 
        & \begin{tabular}{@{}c@{}}9 \\ $d$\end{tabular} 
        & \begin{tabular}{@{}c@{}}10 \\ $D$\end{tabular} 
        & \begin{tabular}{@{}c@{}}11 \\ $n$\end{tabular} 
        & \begin{tabular}{@{}c@{}}12 \\ $N$\end{tabular} \\
        \hline
        $comp_{kh}^{1}$ & 48 & 0 & 10 & 0 & 18 & 31 & 0 & 29 & 0 & 50 & 35 & 34 \\
        \hline
    \end{tabular}  
    \caption{Transformation of a $36$-dimensional octave-specific note-frequency vector into a $12$-dimensional cumulative note-frequency vector.}
    \label{tab:mapping}
\end{table}

\subsection{Proposed Distance Measure}
\label{sec:dis-measure}

After modifying the dataset into a $12$-dimensional cumulative representation, we propose a r\={a}g-aware weighted Euclidean distance to better capture the similarity and dissimilarity among compositions belonging to the same and different r\={a}gs. The proposed distance measure is designed to incorporate the prescribed note structure of a r\={a}g into the distance computation.

Let $a=(a_1,a_2,\dots,a_{12})$ and $b=(b_1,b_2,\dots,b_{12})$ be two $12$-dimensional cumulative note-frequency vectors. Before computing the distance, both vectors are normalized by their total frequency. Thus,
\[
\hat{a}_i
=
\frac{a_i}{\sum_{j=1}^{12} a_j},
\qquad
\hat{b}_i
=
\frac{b_i}{\sum_{j=1}^{12} b_j},
\qquad
i=1,2,\dots,12.
\]

The use of normalized vectors ensures that the distance is not dominated by the total length or total note count of a composition. Instead, the comparison is based on the relative distribution of notes.

The proposed weighted Euclidean distance between two normalized vectors $\hat{a}$ and $\hat{b}$ is defined as
\begin{equation}
\label{eq:weighted-euclidean}
    E_w(a,b)
    =
    \sqrt{
    \sum_{i=1}^{12}
    w_i
    \left(
    \hat{a}_i-\hat{b}_i
    \right)^2
    },
    \quad
    0 < w_i < 1,
    \quad
    \sum_{i=1}^{12} w_i = 1 .
\end{equation}

Here, $w_i$ denotes the weight assigned to the $i$-th note in $\mathcal{N}_{12}$. The weight vector is constructed separately for each r\={a}g according to its prescribed \textit{\={A}roh} and \textit{Avroh} note set.

For each r\={a}g $\rho$, the complete $12$-note set $\mathcal{N}_{12}$ is divided into two mutually exclusive subsets. The first subset, denoted by $\mathcal{N}_{12}^{\prime}$, contains the notes that occur in the prescribed \textit{\={A}roh} and \textit{Avroh} of the corresponding r\={a}g. The second subset, denoted by $\mathcal{N}_{12}^{\prime\prime}$, contains the remaining notes that do not occur in the prescribed \textit{\={A}roh} and \textit{Avroh}. Therefore,
\begin{equation}
    \mathcal{N}_{12}^{\prime}
    \cup
    \mathcal{N}_{12}^{\prime\prime}
    =
    \mathcal{N}_{12},
    \quad
    \mathcal{N}_{12}^{\prime}
    \cap
    \mathcal{N}_{12}^{\prime\prime}
    =
    \emptyset .
\end{equation}

Similarly, let $\mathcal{I}_{\rho}^{\prime}$ denote the set of note indices corresponding to the notes present in the prescribed \textit{\={A}roh} and \textit{Avroh} of r\={a}g $\rho$. Let $\mathcal{I}_{\rho}^{\prime\prime}$ denote the set of remaining note indices. Thus,
\begin{equation}
    \mathcal{I}_{\rho}^{\prime}
    \cup
    \mathcal{I}_{\rho}^{\prime\prime}
    =
    \{1,2,\dots,12\},
    \quad
    \mathcal{I}_{\rho}^{\prime}
    \cap
    \mathcal{I}_{\rho}^{\prime\prime}
    =
    \emptyset .
\end{equation}

In this work, we use a $90:10$ weight distribution. Under this rule, $90\%$ of the total weight is distributed equally among the notes belonging to $\mathcal{I}_{\rho}^{\prime}$, while the remaining $10\%$ of the total weight is distributed equally among the notes belonging to $\mathcal{I}_{\rho}^{\prime\prime}$. Therefore, the weight $w_i$ is defined as
\begin{equation}
    w_i =
    \begin{cases}
    \dfrac{0.90}{|\mathcal{I}_{\rho}^{\prime}|},
    & i \in \mathcal{I}_{\rho}^{\prime},\\[8pt]
    \dfrac{0.10}{|\mathcal{I}_{\rho}^{\prime\prime}|},
    & i \in \mathcal{I}_{\rho}^{\prime\prime}.
    \end{cases}
\end{equation}

This construction ensures that
\[
\sum_{i=1}^{12} w_i = 1.
\]

The purpose of this weighting scheme is to assign greater importance to the notes that define the melodic identity of a r\={a}g. Notes that are part of the prescribed \textit{\={A}roh} and \textit{Avroh} receive higher weights, whereas notes outside the prescribed note set receive lower weights. As a result, the distance computation becomes more sensitive to r\={a}g-specific note usage.

For example, for r\={a}g \textit{Khamaj}, the prescribed note-index set is
\[
\mathcal{I}_{kh}^{\prime}
=
\{1,3,5,6,8,10,11,12\}.
\]
Therefore,
\[
|\mathcal{I}_{kh}^{\prime}|=8,
\qquad
|\mathcal{I}_{kh}^{\prime\prime}|=4.
\]
Hence, under the $90:10$ rule, each prescribed Khamaj note receives weight
\[
\frac{0.90}{8}=0.1125,
\]
whereas each non-prescribed note receives weight
\[
\frac{0.10}{4}=0.0250.
\]

The r\={a}g-specific weight vectors used in this work are presented in Table~\ref{tab:weight-vector}.

\begin{table}[!ht]
    \centering
    \small
    \begin{tabular}{|c|c|c|c|c|c|c|c|c|c|c|c|c|}
        \hline
        \textbf{R\={a}g} 
        & $w_{1}$ & $w_{2}$ & $w_{3}$ & $w_{4}$ & $w_{5}$ & $w_{6}$ 
        & $w_{7}$ & $w_{8}$ & $w_{9}$ & $w_{10}$ & $w_{11}$ & $w_{12}$ \\
        \hline

        \textbf{Bhairavi} 
        & 0.1286 & 0.1286 & 0.0200 & 0.1286 & 0.0200 & 0.1286 
        & 0.0200 & 0.1286 & 0.1286 & 0.0200 & 0.1286 & 0.0200 \\
        \hline


        \textbf{Khamaj} 
        & 0.1125 & 0.0250 & 0.1125 & 0.0250 & 0.1125 & 0.1125
        & 0.0250 & 0.1125 & 0.0250 & 0.1125 & 0.1125 & 0.1125 \\
        \hline
    \end{tabular}
    \caption{R\={a}g-specific weight vectors using the $90:10$ distribution. Here, $90\%$ of the total weight is distributed equally among the notes occurring in the prescribed \textit{\={A}roh} and \textit{Avroh} of the corresponding r\={a}g, while the remaining $10\%$ is distributed equally among the other notes.}
    \label{tab:weight-vector}
\end{table}

\section{Resolving the Observed Shortcomings Using the Modified Distance}
\label{sec:resolving-shortcomings}

The previous examples show that cosine similarity and ordinary Euclidean distance may produce musically misleading results in r\={a}g-based composition analysis. To address these limitations, we revisit the same examples using the proposed modified distance measure. In this approach, each original $36$-dimensional octave-specific note-frequency vector is first converted into a $12$-dimensional cumulative note-frequency vector. The resulting vector is then normalized by total frequency before computing the r\={a}g-aware weighted distance.



The use of the $12$-dimensional cumulative representation reduces the effect of octave-specific note placement. Normalization removes the effect of composition length, while the r\={a}g-aware weight vector gives higher importance to musically significant note positions. The following subsections show how the modified distance addresses the three shortcomings discussed earlier.

\subsection{Case 1: Pure and Non-pure Compositions of the Same R\={a}g}
\label{sec:case1-modified-distance}

In the first case, we compare a pure Khamaj composition, a non-pure Khamaj composition, and a pure Bhairavi composition. The pure Khamaj composition is denoted by $comp_{kh(p)}^{22}$, the non-pure Khamaj composition is denoted by $comp_{kh(np)}^{35}$, and the pure Bhairavi composition is denoted by $comp_{bh(p)}^{43}$.

Earlier, normalized ordinary Euclidean distance on the $36$-dimensional representation incorrectly placed the pure Khamaj composition closer to the pure Bhairavi composition than to the non-pure Khamaj composition. Using the proposed modified distance on the $12$-dimensional cumulative normalized vectors gives the results shown in Table~\ref{tab:case1-modified-distance}.

\begin{table}[!ht]
\centering
\small
\begin{tabular}{|c|c|c|}
\hline
\textbf{Pair of Compositions} & \textbf{Modified Weighted Distance} & \textbf{Interpretation} \\
\hline
$comp_{kh(p)}^{22}$ vs. $comp_{kh(np)}^{35}$ 
& 0.0830 
& Closest pair \\
\hline
$comp_{kh(p)}^{22}$ vs. $comp_{bh(p)}^{43}$ 
& 0.1153 
& Farther than Khamaj pair \\
\hline
\end{tabular}
\caption{Resolution of the first shortcoming using the modified weighted distance. The pure and non-pure Khamaj compositions remain closer to each other than to the pure Bhairavi composition.}
\label{tab:case1-modified-distance}
\end{table}

From Table~\ref{tab:case1-modified-distance}, we observe that
\[
D_w(comp_{kh(p)}^{22},comp_{kh(np)}^{35})
<
D_w(comp_{kh(p)}^{22},comp_{bh(p)}^{43})
\]
Thus, the modified weighted distance gives the expected musical ordering by keeping the pure and non-pure Khamaj compositions closest to each other.

\subsection{Case 2: Octave-wise Non-overlapping Notes}
\label{sec:case2-modified-distance}

In the second case, we revisit the two Bhairavi compositions whose non-zero note indices were distributed across different octaves in a non-overlapping manner. In the original $36$-dimensional representation, the cosine similarity between these two compositions became zero because there was no overlap in their note-index positions.

However, from a r\={a}g-level perspective, the octave position of the same note does not change the identity of the r\={a}g. Therefore, the proposed $12$-dimensional cumulative representation combines corresponding notes across the three octaves before computing the distance. The modified distance result is shown in Table~\ref{tab:case2-modified-distance}.

\begin{table}[!ht]
\centering
\small
\begin{tabular}{|c|c|c|}
\hline
\textbf{Pair of Compositions} & \textbf{Modified Weighted Distance} & \textbf{Interpretation} \\
\hline
$comp_{bh(np)}^{312}$ vs. $comp_{bh(np)}^{415}$ 
& 0.0647 
& Small distance after octave aggregation \\
\hline
\end{tabular}
\caption{Resolution of the second shortcoming using the modified weighted distance. Although the two Bhairavi compositions had non-overlapping note indices in the $36$-dimensional representation, the octave-summed representation shows that they are close.}
\label{tab:case2-modified-distance}
\end{table}

Table~\ref{tab:case2-modified-distance} shows that the modified distance between the two Bhairavi compositions is small. Thus, the proposed method removes the artificial dissimilarity caused by octave-wise non-overlap. This resolves the earlier problem where cosine similarity became zero despite both compositions belonging to the same r\={a}g.

\subsection{Case 3: Pure Compositions Belonging to Different R\={a}gs}
\label{sec:case3-modified-distance}

In the third case, we revisit the example where cosine similarity gave a higher score to two pure compositions belonging to different r\={a}gs than to two pure compositions belonging to the same r\={a}g. Specifically, the cosine similarity between $comp_{bh(p)}^{43}$ and $comp_{kh(p)}^{17}$ was higher than the cosine similarity between $comp_{kh(p)}^{20}$ and $comp_{kh(p)}^{21}$.

Using the proposed modified distance on the same examples, we obtain the results shown in Table~\ref{tab:case3-modified-distance}.

\begin{table}[!ht]
\centering
\small
\begin{tabular}{|c|c|c|}
\hline
\textbf{Pair of Compositions} & \textbf{Modified Weighted Distance} & \textbf{Interpretation} \\
\hline
$comp_{kh(p)}^{20}$ vs. $comp_{kh(p)}^{21}$ 
& 0.1000 
& Same r\={a}g pair is closer \\
\hline
$comp_{bh(p)}^{43}$ vs. $comp_{kh(p)}^{17}$ 
& 0.1062 
& Different r\={a}g pair is farther \\
\hline
\end{tabular}
\caption{Resolution of the third shortcoming using the modified weighted distance. The two pure Khamaj compositions are assigned a smaller distance than the pure Bhairavi--pure Khamaj pair.}
\label{tab:case3-modified-distance}
\end{table}

From Table~\ref{tab:case3-modified-distance}, we observe that
\[
D_w(comp_{kh(p)}^{20},comp_{kh(p)}^{21})
<
D_w(comp_{bh(p)}^{43},comp_{kh(p)}^{17}).
\]
This is the musically expected ordering, since two pure compositions belonging to the same r\={a}g should be closer than two pure compositions belonging to different r\={a}gs.

Overall, these three cases demonstrate that the proposed modified distance addresses the shortcomings of cosine similarity and ordinary Euclidean distance. The octave-summed representation reduces octave-position dependency, normalization reduces the effect of composition length, and the r\={a}g-aware weighting scheme emphasizes musically important note positions.

In Section~\ref{sec:perf-knn}, we demonstrate the effectiveness of the proposed weighted distance measure by evaluating the modified dataset using a $k$-nearest-neighbor classifier.

\subsection{R\={a}g Prediction Evaluation using K-Nearest-Neighbour Classifier}
\label{sec:perf-knn}

In this section, we evaluate the effectiveness of the proposed distance measure for r\={a}g prediction using the $k$-nearest-neighbour ($k$NN) classifier. We compare the classification accuracy obtained using ordinary Euclidean distance and the proposed weighted Euclidean distance for different values of $k$.

For this experiment, we select compositions belonging to three r\={a}gs, namely \textit{Bhairavi}, \textit{Bihag}, and \textit{Khamaj}. These are among the most frequently occurring r\={a}gs in the dataset. Although \textit{Kirtan} is also one of the most frequent labels, it is excluded from this experiment because it does not correspond to a specific Hindustani classical r\={a}g and therefore does not follow a fixed \textit{\={A}roh} and \textit{Avroh} structure. Thus, the classification task is formulated as a three-class r\={a}g prediction problem using $186$ compositions represented by $12$ cumulative note-frequency features.

The dataset is divided into a training set and a test set using a $75:25$ split. The training set is used to build the $k$NN classifier, while the test set is used to compute the prediction accuracy. For the weighted Euclidean distance, we consider two weight distributions, namely $90:10$ and $80:20$. In each case, r\={a}g-specific weight vectors are constructed using the prescribed \textit{\={A}roh} and \textit{Avroh} note sets of the selected r\={a}gs, as shown in Table~\ref{tab:weight-vector}.

Table~\ref{tab:accuracy-score} presents the accuracy scores obtained for different values of $k$ using ordinary Euclidean distance and the proposed weighted Euclidean distance.

\begin{table}[!ht]
\small
\centering
\begin{tabular}{|c|c|c|c|c|}
\hline
\multirow{2}{*}{\textbf{$k$}} &
\multirow{2}{*}{\begin{tabular}{@{}c@{}}\textbf{Euclidean}\\\textbf{Acc.}\end{tabular}} &
\multirow{2}{*}{\textbf{R\={a}g}} &
\multicolumn{2}{c|}{\begin{tabular}{@{}c@{}}\textbf{Weighted Euclidean}\\\textbf{($N_r\!:\!N_{nr}$)}\end{tabular}} \\
\hhline{|~|~|~|--|}
& & & \textbf{90:10} & \textbf{80:20} \\
\hline

\multirow{3}{*}{3} & & Bhairavi & 0.7447 & 0.7021 \\
\hhline{|~|~|---|}
& 0.7660 & Bihag & \textbf{0.7872} & \textbf{0.7660} \\
\hhline{|~|~|---|}
& & Khamaj & 0.7234 & \textbf{0.7660} \\
\hline

\multirow{3}{*}{4} & & Bhairavi & 0.7660 & 0.7660 \\
\hhline{|~|~|---|}
& \cellcolor{blue!25}0.7872 & Bihag & \cellcolor{blue!25}\textbf{0.8511} & \textbf{0.8298} \\
\hhline{|~|~|---|}
& & Khamaj & 0.7660 & 0.7872 \\
\hline

\multirow{3}{*}{5} & & Bhairavi & 0.6383 & 0.7447 \\
\hhline{|~|~|---|}
& 0.7660 & Bihag & \textbf{0.8085} & \textbf{0.8085} \\
\hhline{|~|~|---|}
& & Khamaj & 0.7660 & 0.7872 \\
\hline

\multirow{3}{*}{6} & & Bhairavi & 0.7021 & 0.7446 \\
\hhline{|~|~|---|}
& 0.7446 & Bihag & \textbf{0.8085} & \cellcolor{blue!25}\textbf{0.8510} \\
\hhline{|~|~|---|}
& & Khamaj & 0.7234 & 0.8290 \\
\hline

\multirow{3}{*}{7} & & Bhairavi & 0.6600 & 0.6600 \\
\hhline{|~|~|---|}
& 0.7021 & Bihag & 0.7021 & 0.7660 \\
\hhline{|~|~|---|}
& & Khamaj & \textbf{0.7447} & \textbf{0.8085} \\
\hline
\end{tabular}
\caption{Comparison of $k$NN accuracy scores for different values of $k$ using ordinary Euclidean distance and weighted Euclidean distance.}
\label{tab:accuracy-score}
\end{table}

For each value of $k$, the Euclidean accuracy is obtained using the ordinary Euclidean distance on the modified $12$-dimensional representation. For the weighted Euclidean distance, we evaluate the classifier using r\={a}g-specific weight vectors and two different weight distributions, namely $90:10$ and $80:20$. The best accuracy values obtained among the weighted-distance settings are shown in bold.

The motivation for using r\={a}g-specific weight vectors is that the importance of a note depends on the r\={a}g under consideration. A note that is musically important for one r\={a}g may not be equally important for another r\={a}g. Therefore, while computing the weighted distance between a test composition and the training compositions, the distance is evaluated with respect to the weight vectors of the candidate r\={a}gs. The nearest neighbours are then identified based on the resulting weighted distances, and the r\={a}g label is predicted using majority voting among the $k$ nearest neighbours.

From Table~\ref{tab:accuracy-score}, it can be observed that the proposed weighted Euclidean distance generally improves the $k$NN classification accuracy compared to ordinary Euclidean distance. For example, when $k=4$, the ordinary Euclidean distance gives an accuracy of $0.7872$, whereas the weighted Euclidean distance with the $90:10$ distribution and the Bihag weight vector gives the highest accuracy of $0.8511$. Similarly, for $k=6$, the weighted Euclidean distance with the $80:20$ distribution achieves an accuracy of $0.8510$, which is higher than the corresponding Euclidean accuracy of $0.7446$.

These results indicate that incorporating r\={a}g-specific note importance into the distance computation improves the discriminative ability of the $k$NN classifier. Thus, the proposed weighted Euclidean distance is more suitable than ordinary Euclidean distance for r\={a}g prediction using symbolic note-frequency representations.

\section{Conclusions and Future Work}
\label{sec:conclusion}

In this paper, we presented a symbolic music-based approach for r\={a}g classification of Rabindra Sangeet compositions. The study was motivated by the fact that Rabindra Sangeet often draws upon Hindustani r\={a}g structures while allowing considerable creative freedom in melodic treatment. This makes automatic r\={a}g identification challenging, especially when standard similarity and distance measures are directly applied to note-frequency representations.

To address this problem, we prepared a supervised symbolic dataset of r\={a}g-labelled Tagore songs from \textit{Swarabitan}. Each composition was represented through its note-frequency distribution, initially over $36$ notes spanning three octaves. The experimental analysis showed that conventional cosine similarity and Euclidean distance may sometimes produce misleading similarity scores. In particular, compositions belonging to the same r\={a}g may appear dissimilar due to octave-wise non-overlapping note positions, while compositions belonging to different r\={a}gs may appear highly similar because of comparable note-frequency patterns.

To overcome these limitations, we modified the feature representation by aggregating notes across octaves into a $12$-note representation. Furthermore, we proposed a \textit{Weighted Euclidean Distance} measure that assigns higher importance to notes belonging to the characteristic \={A}roh and Avroh of a r\={a}g, and lower importance to the remaining notes. The proposed distance measure provides a musically informed way of comparing compositions, as it incorporates r\={a}g-specific note importance rather than treating all note-frequency differences equally. The usefulness of the proposed measure was demonstrated through its application in a $k$-nearest-neighbor classifier, where it helped improve the distinction between compositions belonging to the same and different r\={a}gs.

The present work opens several directions for future research. First, the dataset can be expanded by including a larger number of compositions from the complete \textit{Swarabitan} collection, so that r\={a}gs with fewer available samples can also be studied more effectively. Second, the weight values used in the proposed distance measure can be learned automatically from data instead of being assigned manually. This may help obtain more adaptive and r\={a}g-specific weighting schemes. Third, future work may include sequential melodic features such as note transitions, phrase patterns, and Markov or hidden Markov model-based representations, since r\={a}g identity is often expressed not only through note frequency but also through melodic movement. 

In addition, the proposed symbolic framework can be extended by incorporating other musical attributes such as t\={a}l, phrase structure, ornamentation, and melodic motifs. Finally, the method can be compared with other machine learning and deep learning models, as well as with audio-based r\={a}g recognition approaches, to develop a more comprehensive framework for computational analysis of Rabindra Sangeet.

\bibliographystyle{ieeetr}
\bibliography{bibliography}
\end{document}